\documentclass[twocolumn]{aastex631}
\usepackage{amsmath}
\DeclareMathOperator\arctanh{arctanh}
\DeclareMathOperator\Cov{Cov}

\graphicspath{{./}{plots/}}

\begin{document}

\title{Indefinitely Flat Circular Velocities and the Baryonic Tully-Fisher Relation from Weak Lensing}

\author[0000-0001-7048-3173]{Tobias Mistele}
\affiliation{Department of Astronomy, Case Western Reserve University, 10900 Euclid Avenue, Cleveland, Ohio 44106, USA}

\author[0000-0002-9762-0980]{Stacy McGaugh}
\affiliation{Department of Astronomy, Case Western Reserve University, 10900 Euclid Avenue, Cleveland, Ohio 44106, USA}

\author[0000-0002-9024-9883]{Federico Lelli}
\affiliation{INAF — Arcetri Astrophysical Observatory, Largo Enrico Fermi 5, 50125 Firenze, Italy}

\author[0000-0003-2022-1911]{James Schombert}
\affiliation{Department of Physics, University of Oregon, 1371 E 13th Ave, Eugene, Oregon 97403, USA}

\author[0000-0002-6707-2581]{Pengfei Li}
\affiliation{School of Astronomy and Space Science, Nanjing University, Nanjing, Jiangsu 210023, China}

\begin{abstract}
We use a new deprojection formula to infer the gravitational potential around isolated galaxies from weak gravitational lensing. The results imply circular velocity curves that remain flat for hundreds of kpc, greatly extending the classic result from 21 cm observations. Indeed, there is no clear hint of a decline out to 1 Mpc, well beyond the expected virial radii of dark matter halos. Binning the data by mass reveals a correlation with  the flat circular speed that closely agrees with the Baryonic Tully-Fisher Relation known from kinematic data. These results apply to both early and late type galaxies, indicating a common universal behavior.
\end{abstract}

\section{Introduction}

The rotation curves of spiral galaxies become approximately flat at large radii \citep{Rubin1978} and remain so well beyond the extent of the observed luminous mass \citep{Bosma1981}. A constant rotation speed implies an enclosed mass that increases linearly without bound, a behavior that should not persist indefinitely.
A long-standing question is just how far it does persist.

Rotation curves are commonly inferred from radio interferometry of the 21 cm line of atomic hydrogen.
This technique allows probing many tens of $\mathrm{kpc}$ \citep{Lelli2016} and sometimes up to $100\,\mathrm{kpc}$ \citep{Noordermeer2005,Lelli2010}, without revealing any credible indication of a Keplerian decline \citep{deBlok2008,Lelli2022}. Rotation curves have distinctive shapes that correlate with surface brightness \citep{Lelli2013, Lelli2016CDR} and are not \textit{perfectly} flat \citep{Casertano1991}: it is common to see a gradual decline in massive galaxies, but the gradient is not Keplerian and the rotation curve tends to flatten out at the largest probed radii \citep{Noordermeer2005, DiTeodoro2023}.
 
Weak gravitational lensing offers another probe \citep[e.g.,][]{Hudson1998,Kleinheinrich2006,Brimioulle2013,Milgrom2013,Wang2016} that extends to much larger radii. Indeed, recent lensing observations imply rotation curves that remain remarkably flat to a few hundred $\mathrm{kpc}$ \citep{Brouwer2021}.
Here we apply a new technique \citep{Mistele2023d} to further extend these results.

Galaxies follow the Baryonic Tully-Fisher relation (BTFR), which links the  baryonic mass (stars plus gas) to the mean rotation speed along the flat part of the rotation curve \citep{McGaugh2000,Lelli2016c,Schombert2020}.
The BTFR generalizes the original Tully-Fisher relation \citep{Tully1977}, which relates luminosity and line-width as proxies for stellar mass and rotation speed \citep{Verheijen2001,Ponomareva2017,Ponomareva2018,Lelli2019}.

Here, we derive circular velocity curves and the corresponding BTFR from weak lensing data.
We measure circular velocities out to $\sim 1\,\mathrm{Mpc}$, exploiting the KiDS DR4 weak-lensing data \citep{Kuijken2019,Giblin2021,Bilicki2021}, that was previously analyzed in \citet{Brouwer2021}, but using a  new robust deprojection method from \citet{Mistele2023d}.

\section{Data}
\label{sec:data}

Inspired by \citet{Brouwer2021}, we analyze a sample of isolated galaxies from the KiDS survey \citep{Kuijken2019}.
For distance dependent quantities, we adopt a Hubble constant $H_0 = 73\,\mathrm{km}\,\mathrm{s}^{-1}\,\mathrm{Mpc}^{-1}$.
This choice is made for consistency with previous kinematic work \citep{Lelli2016} and $H_0$ measured with the BTFR \citep{Schombert2020}.
We further assume a flat FLRW cosmology with $\Omega_{\mathrm{m}} = 0.2793$ \citep{Hinshaw2013} for consistency with \citet{Brouwer2021}.
Where useful, we employ the notation $h_{70} \equiv H_0/(70\,\mathrm{km}\,\mathrm{s}^{-1}\,\mathrm{Mpc}^{-1})$.

We follow the procedure of \citet{Mistele2023d} to select lens and source galaxy samples.
Specifically, we use source galaxies from the KiDS-1000 SOM-gold catalog \citep{Kuijken2019,Wright2020,Giblin2021,Hildebrandt2021} and lens galaxies from the KiDS-bright sample \citep{Bilicki2021}.
We split the lens sample into four baryonic mass bins.
To have a good compromise between bin width and number of galaxies in each bin, we use bin edges $\log_{10} M_b/M_\odot = [9.0, 10.5, 10.8, 11.1, 11.5]$.
Unlike \citet{Mistele2023d} we do not impose an explicit cutoff on stellar mass.

We calculate stellar and baryonic masses as in \citet{Mistele2023d} where we reanalyzed the KiDS data using the stellar population synthesis model of \citet{Schombert2014}, which was previously used for the kinematic BTFR in \citet{Lelli2019}.
Compared to \citet{Brouwer2021}, we find excellent agreement for late-type galaxies (LTGs) but slightly larger stellar masses for early-type galaxies (ETGs).
Consequently, we adopt the stellar masses of \citet{Brouwer2021} for LTGs and correct the stellar masses of ETGs by a factor of $1.4$.
Following \citet{Mistele2023d} and \citet{Brouwer2021} we define LTGs and ETGs by the color split $u-r \gtrless 2.5$.
We correct all stellar masses to account for our choice of $H_0$. 

We use scaling relations to account for the baryonic mass in gas.
For ETGs, we add a hot gas component according to the scaling relation
\begin{align}
 \label{eq:fhot}
 \frac{M_{g,\mathrm{hot}}}{M_\ast} = 10^{-5.414} \cdot \left(\frac{M_\ast}{M_\odot}\right)^{0.47} \,,
\end{align}
which accounts for the X-ray-emitting coronae of massive ETGs \citep{Chae2021}.

Star-forming LTGs have a non-negligible interstellar medium of atomic and molecular gas.
Thus, for LTGs we add a cold gas component according to the scaling relation
\begin{align}
 \label{eq:fcold}
 \frac{M_{g,\mathrm{cold}}}{M_\ast} = \frac{1}{X} \left(11550 \left(\frac{M_\ast}{M_\odot}\right)^{-0.46} + 0.07 \right) \,,
\end{align}
where
\begin{align}
 \label{eq:metal}
 X = 0.75  - 38.2 \left(\frac{M_\ast}{1.5\cdot10^{24}M_\odot}\right)^{0.22} \,.
\end{align}
The first term in Equation~\eqref{eq:fcold} represents atomic gas according to the scaling relation from \citet{Lelli2016} with $M_\ast = 0.5 L_{[3.6]}$.
The second term takes into account molecular gas.
Equation~\eqref{eq:metal} accounts for the variation of the hydrogen fraction $X$ as metallicity varies with stellar mass \citep{McGaugh2020b}.

Since we are interested in the gravitational potential of isolated galaxies, we require that lens galaxies have no neighbor with a fraction $f_\star>0.1$ of their stellar mass within a 3D distance $R_{\mathrm{isol}} = 4\,\mathrm{Mpc}/h_{70}$.\footnote{
 Following \citet{Mistele2023d}, we adopt the stellar masses of \citet{Brouwer2021} when imposing this isolation criterion. Compared to using our own stellar masses, this makes the isolation criterion for ETGs more strict than for LTGs which helps counter the fact that ETGs are more clustered \citep{Dressler1980}.}
As shown by \citet{Mistele2023d}, the weaker isolation criterion $R_{\mathrm{isol}} = 3\,\mathrm{Mpc}/h_{70}$ used by \citet{Brouwer2021} is not sufficient for ETGs, likely because these galaxies are more clustered than LTGs \citep{Dressler1980}.

As discussed in \citet{Mistele2023d} and \citet{Brouwer2021}, the isolation criterion relies on the KiDS photometric redshifts, which have significant uncertainties. Using $\Lambda$CDM simulations, \citet{Brouwer2021} estimate that their weaker isolation criterion is reliable out to about $300\,\mathrm{kpc}$. 
Empirically, we find that our overall results do not change out to $\sim1\,\mathrm{Mpc}$ for any sensible choice of $f_\star$ and $R_{\mathrm{isol}}$, so our results may be reliable out to radii $\sim$3 times larger than what \citet{Brouwer2021} estimate. In the following, we consider radii out to $1\,\mathrm{Mpc}$ but, if in doubt, $300\,\mathrm{kpc}$ is a conservative lower bound. 
In addition, to estimate the impact of redshift uncertainties, we compared to a sample of lenses with spectroscopic redshifts from GAMA~III \citep{Driver2022,Bellstedt2020,Taylor2011} and found consistent results (see Section~\ref{sec:results:vc}).

The azimuthally averaged tangential shear that we use \citep{Mistele2023d} is a measure of the cumulative projected mass along the line-of-sight \citep[e.g.][]{Kaiser1993}, so an isolation criterion based on a cylinder along the line-of-sight instead of a 3D sphere may, in principle, be preferable.
However, as discussed in \citet{Mistele2023d}, this may leave too few lenses to obtain a useful signal.
This will be investigated in 
detail in future work.

After applying our isolation criterion, the four mass bins contain $[24297, 17183, 22083, 11570]$ LTG lenses and $[1795, 5439, 14220, 32440]$ ETG lenses.
The lowest mass bin for ETGs has too few lenses to give any useful signal, so we do not consider it further.
The isolation criterion leaves a fraction of $8\%$, $18\%$, $30\%$, and $42\%$ of LTG lenses and $9\%$, $14\%$, and $22\%$ of ETG lenses, respectively, in each mass bin. Not split by mass and type, the isolated fraction is $16\%$.

\section{Method}
\label{sec:method}

\subsection{Circular velocities}
\label{sec:method:vc}

The weak-lensing signal of any individual lens is indeed weak, so we stack all lenses in a given mass bin.
In particular, we derive stacked circular velocities using the method from \citet{Mistele2023d} to infer stacked radial accelerations, $g_{\mathrm{obs}}^{\mathrm{stacked}}(R)$, in 15 logarithmic bins between $0.3\,\mathrm{Mpc}/h_{70}$ and $3\,\mathrm{Mpc}/h_{70}$.
These stacked radial accelerations are weighted averages of the radial accelerations $g_{\mathrm{obs},l}(R)$ of each lens $l$,
\begin{align}
 \label{eq:gobs_stacked}
 g_{\mathrm{obs}}^{\mathrm{stacked}}(R) = \bar{N}^{-1}(R) \sum_l \bar{w}_l(R) \, g_{\mathrm{obs},l}(R) \,,
\end{align} 
with weights $\bar{w}_l(R)$ and the normalization factor $\bar{N}^{-1}(R) = \sum_l \bar{w}_l(R)$.
Following \citet{Mistele2023d}, we choose the weights $\bar{w}_l(R)$ to be the inverse square of the statistical uncertainty of $g_{\mathrm{obs},l}(R)$, i.e. $\bar{w}_l(R) = \sigma^{-2}_{g_{\mathrm{obs},l}}(R)$.
The $g_{\mathrm{obs},l}(R)$ of each lens $l$ is calculated from the excess surface density (ESD) profile $\Delta \Sigma_l(R)$ of that lens using the deprojection formula \citep{Mistele2023d},
\begin{align}
 \label{eq:gobs_from_esd}
 g_{\mathrm{obs}}(R) = \frac{G M(R)}{R^2} =  4G \int_0^{\pi/2} d\theta \, \Delta \Sigma\left(\frac{R}{\sin \theta}\right) \,,
\end{align}
where $M(R)$ is the deprojected dynamical mass enclosed within a spherical radius $R$. 
This deprojection formula assumes spherical symmetry which is a reasonable approximation at the large radii we consider in this work.

The integral in Equation~\eqref{eq:gobs_from_esd} is evaluated as in \citet{Mistele2023d}.
In particular, the systematic uncertainties on $g_{\mathrm{obs}}(R)$ are calculated considering different choices in how to interpolate between the discrete $\Delta \Sigma(R)$ data points and how to extrapolate $\Delta \Sigma(R)$ beyond the last data point at $R=R_{\mathrm{max}}$ (see \citealt{Mistele2023d} for technical details).
These systematic uncertainties become important only close to $R_{\mathrm{max}}$.
In practice, this happens around $1\,\mathrm{Mpc}$ (see Section~\ref{sec:results:vc}).
For the statistical uncertainties, we adopt an improved procedure described in Appendix~\ref{sec:appendix:staterrors}.

The stacked radial accelerations $g_{\mathrm{obs}}^{\mathrm{stacked}}(R)$ can be converted to stacked circular velocities using
\begin{align}
\label{eq:vc}
V_c(R) \equiv (R \, g_{\mathrm{obs}}^{\mathrm{stacked}}(R))^{1/2} \,.
\end{align}
These $V_c$ values are not linear averages of the circular velocities of the individual stacked lens galaxies, i.e., they are not $\langle V_c \rangle$.
Instead, they are $\sqrt{\langle V_c^2 \rangle}$.

\begin{figure*}
\plotone{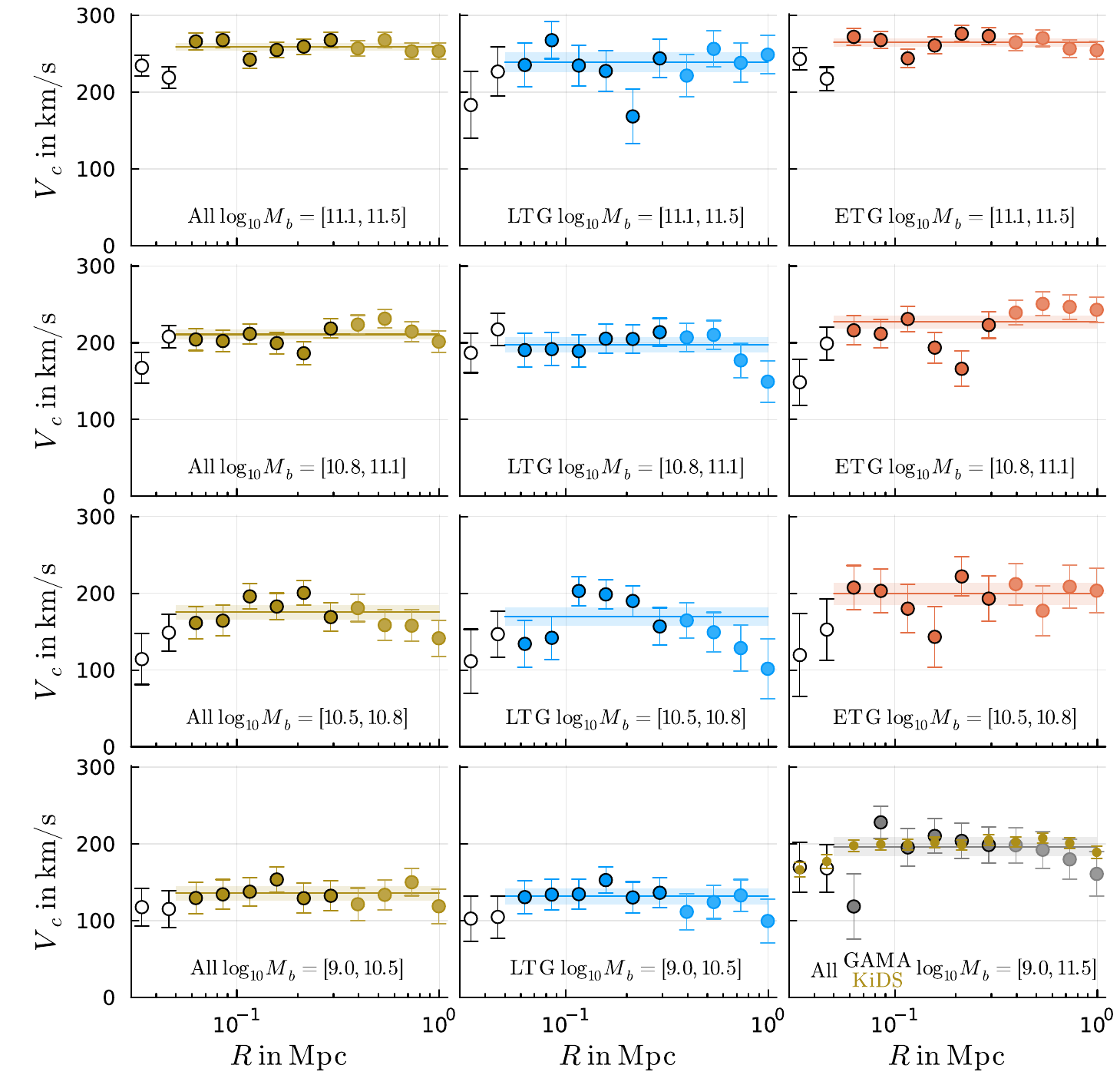}  
 \caption{
  Circular velocities implied by weak lensing for four baryonic mass bins (most to least massive from the top row to the bottom) for the whole sample (left column), for LTGs (middle column), and for ETGs (right column).
  The lowest ETG mass bin is not shown because it contains too few lenses.
  Instead we show results for lenses with spectroscopic redshifts from GAMA, without splitting by mass or type due to the small sample size (gray and white symbols).
  For comparison, we also show results for KiDS without splitting by mass or type (small yellow symbols).
  Open symbols at small radii indicate where lenses are not yet effective point masses.
  Light-colored symbols at large radii indicate data points that may still be reliable but where the isolation criterion is less certain.
  The error bars show the statistical errors.
  Horizontal lines and the corresponding shaded regions indicate the inferred $V_{\mathrm{flat}}$ values and uncertainties that we use for the BTFR.
  The extent of the horizontal lines indicates the radial range we consider when calculating $V_{\mathrm{flat}}$.
 }
 \label{fig:vc}
\end{figure*}

\subsection{Averaged quantities for BTFR}
\label{sec:method:vflat}
\label{sec:method:Mb}

\begin{figure*}
\plotone{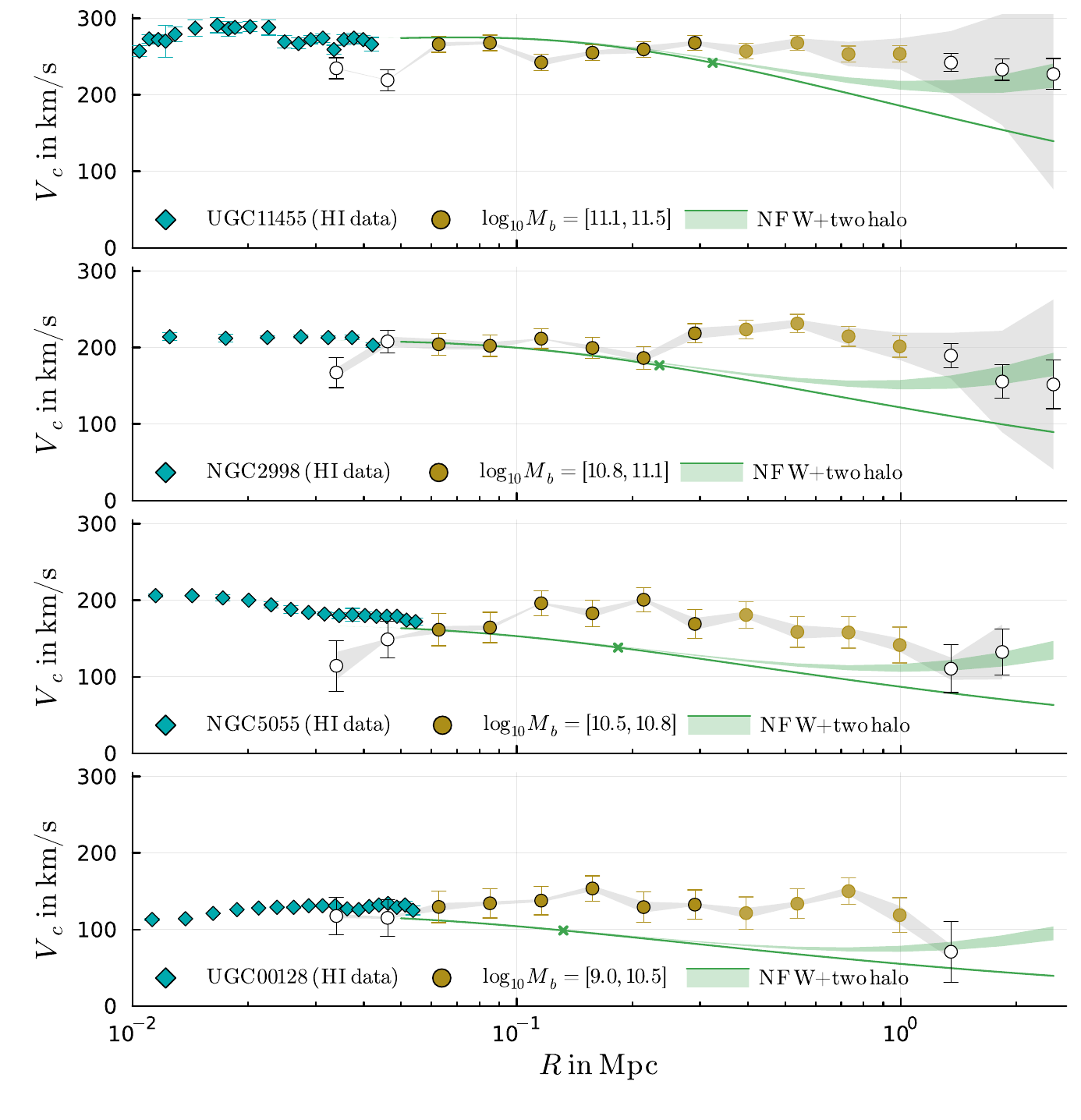}
 \caption{
   The circular velocities from weak lensing (circles) compared with those from gas kinematics (diamonds).
   The individual galaxies illustrated here have among the most extended 21 cm rotation curves in their mass bins; the lensing data continue to much larger radii still. 
   The error bars show the statistical error, while the gray band indicates the systematic uncertainty from converting ESD profiles to radial accelerations (Section~\ref{sec:method:vc}).
  Symbol colors are as in Figure~\ref{fig:vc}.
  Open symbols at large radii indicate where lenses are not sufficiently isolated.
  The solid green lines indicate the circular velocities of NFW halos and baryonic point masses appropriate for each mass bin.
  Green crosses indicate each NFW halo's virial radius.
  The light green band adds a qualitative estimate of a two-halo term contribution to the NFW halo, which may become important at large radii in case our isolation criterion is imperfect there.
  }
 \label{fig:vc-sparc}
\end{figure*}

In Section~\ref{sec:results:vc}, we show that the circular-velocity curves from weak lensing are approximately flat at large radii. Thus, we can construct a BTFR, taking the flat circular velocity $V_{\mathrm{flat}}$ to be a 
weighted average of $V_c(R)$ at different radii $R$.
We choose the weight of each $V_c(R)$ to be proportional to the inverse square of its statistical uncertainty.
We average over the radii $50\,\mathrm{kpc} < R < 1000\,\mathrm{kpc}$, where the lower cutoff is chosen such that the baryonic mass of the lenses can effectively be treated as a point mass and the upper cutoff is chosen to avoid systematic uncertainties (see Section~\ref{sec:method:vc} and Figure~\ref{fig:vc-sparc}) and because the isolation criterion is probably not reliable beyond $1\,\mathrm{Mpc}$.
We consider the effect of other choices in Section~\ref{sec:results:btfr}. 

To construct the BTFR we need both $V_{\mathrm{flat}}$ and $M_b$.
Our $V_{\mathrm{flat}}$ is based on stacked velocities $\sqrt{\langle V_c^2 \rangle}$ inferred from weak lensing.
Thus, to test the BTFR using weak-lensing data, we need a suitably stacked $M_b$ to compare to.
Since the relation $V_c \propto M_b^{1/4}$ \citep{McGaugh2000} concerns individual galaxies, the appropriate quantity to compare the stacked $V_c$ to is $\sqrt{\langle \sqrt{M_b} \rangle}$ and not, for example, $\langle M_b \rangle^{1/4}$.
This averaging procedure has the important property that, if the relation $V_c \propto M_b^{1/4}$ holds for individual galaxies, it also holds for the stacked quantities.
For $V_{\mathrm{flat}}$, one additionally needs to take into account the averaging over radial bins described above, see Appendix~\ref{sec:appendix:btfr} for details.\footnote{
 If $V_{\mathrm{flat}} \propto M_b^n$ follows a power slightly different from $n=1/4$, this is only a very small numerical effect, so for simplicity we consider $n=1/4$ when averaging $M_b$.}

For our lens sample, these properly-averaged baryonic masses $M_{b,\mathrm{eff}}$ are smaller than those obtained from a naive unweighted linear average, but
the difference plays only a minor role. Indeed, the properly and naively averaged masses differ by $<4\%$, except in the lowest mass bin where $M_{b,\mathrm{eff}}$ is about $20\%$ smaller.

\section{Results}

\subsection{Circular velocities}
\label{sec:results:vc}

Figure~\ref{fig:vc} and Table~\ref{tab:vc} show the circular velocities inferred from weak lensing.
We split the data by mass, doing so for the whole sample and for LTGs and ETGs separately.
These circular velocity curves are approximately flat out to $\sim 1\,\mathrm{Mpc}$, with no clear indication of a decline.
This remarkable behavior persists in every mass bin for both ETGs and LTGs.

Since our isolation criterion is stricter than that of \citet{Brouwer2021}, these results may be reliable all the way out to $\sim1\,\mathrm{Mpc}$ (see Section~\ref{sec:data}).
Even conservatively, our results are reliable out to at least $R = 300\,\mathrm{kpc}$, where the weaker isolation criterion of \citet{Brouwer2021} was shown to be reliable using $\Lambda$CDM simulations.

As a cross-check, Figure~\ref{fig:vc} shows the circular velocity curve of a sample of GAMA~III lenses with spectroscopic redshifts.
We analyse these in the same way as the KiDS lenses.
The spectroscopic GAMA sample is much smaller ($13,957$, or $9\%$, isolated lenses) than the KiDS sample with photometric redshifts, so we show only a single wide mass bin for GAMA and not split by type.
We find the same behavior as for KiDS.
GAMA shows a slightly stronger but not very significant decline beyond $\sim500\,\mathrm{kpc}$.

Figure~\ref{fig:vc-sparc} shows how the weak-lensing data compares to typical rotation curves from gas kinematics.\footnote{ A few data points at large radii are missing. For these, $g_{\mathrm{obs}}^{\mathrm{stacked}}$ is negative so that we cannot calculate a circular velocity.}
For illustration, we choose galaxies from the SPARC database \citep{Lelli2016} with comparable $M_b$ and $V_{\mathrm{flat}}$ to each weak-lensing bin:
UGC128 \citep{Verheijen1999}, NGC5055 \citep{BlaisOuellette2004, Battaglia2006}, NGC2998 \citep{Broeils1992}, and UGC11455 \citep{Spekkens2006}.
For reference, these have stellar effective radii between $4\,\mathrm{kpc}$ and 
$10\,\mathrm{kpc}$: over half the stellar mass is encompassed within the first point plotted in Figure~\ref{fig:vc-sparc}.

Weak lensing data extend the circular velocity curves from gas kinematics by more than one order of magnitude in radius.
Rotation curves remain flat to $\sim 1\,\mathrm{Mpc}$.
Beyond $1\,\mathrm{Mpc}$, the circular velocities in some mass bins possibly 
decline, but there is no clear departure from flatness, let alone any indication of a Keplerian decline. At these extreme radii, systematic uncertainties on $V_c$ become significant and the continued isolation of the lenses becomes dubious.

\subsection{Comparison with $\Lambda$CDM expectations}

Figure~\ref{fig:vc-sparc} also shows the rotation curve implied by a dark matter (DM) halo and a baryonic point mass, $V_c^2(R) = V_{c,\mathrm{DM}}^2(R) + G M_{b,\mathrm{eff}}/R$ with the averaged baryonic mass $M_{b,\mathrm{eff}}$ (Section~\ref{sec:method:Mb}).
For simplicity, we assume an NFW halo \citep{Navarro1996}.
The halo parameters are determined using the WMAP5 mass-concentration relation from \citet{Maccio2008} and the stellar mass-halo mass relation from \citet{Kravtsov2018}, assuming the average stellar mass in each bin.
The \citet{Kravtsov2018} relation is the most appropriate because, unlike other common relations, it does not overshoot the circular velocities at small radii in the high-mass bins \citep[][]{DiCintio2016,Li2022a}.

\begin{deluxetable*}{lCCCCC}
 \caption{Circular Velocities from Lensing \label{tab:vc}}
 \tablehead{%
		  \colhead{Sample}
		& \colhead{$\log_{10} R$}
		 & \colhead{$V_c$ (bin $1$)} & \colhead{$V_c$ (bin $2$)} & \colhead{$V_c$ (bin $3$)} & \colhead{$V_c$ (bin $4$)}%
		\\
		   \colhead{}
		&  \colhead{$\mathrm{kpc}$}
		 & \colhead{$\mathrm{km/s}$} & \colhead{$\mathrm{km/s}$} & \colhead{$\mathrm{km/s}$} & \colhead{$\mathrm{km/s}$}}\startdata 
			  All
			& $1.53$%
						& $
							117.6
							\pm
							24.7
							\pm
							2.8
						$
						& $
							114.3
							\pm
							33.1
							\pm
							18.1
						$
						& $
							167.3
							\pm
							19.7
							\pm
							5.8
						$
						& $
							234.6
							\pm
							13.7
							\pm
							0.6
						$
					\\
			  All
			& $1.66$%
						& $
							115.2
							\pm
							23.9
							\pm
							1.2
						$
						& $
							148.8
							\pm
							23.9
							\pm
							0.7
						$
						& $
							207.9
							\pm
							14.9
							\pm
							6.2
						$
						& $
							219.0
							\pm
							13.6
							\pm
							0.2
						$
					\\
			  All
			& $1.80$%
						& $
							129.5
							\pm
							20.5
							\pm
							5.0
						$
						& $
							161.4
							\pm
							21.0
							\pm
							4.8
						$
						& $
							204.2
							\pm
							14.4
							\pm
							6.9
						$
						& $
							266.0
							\pm
							10.6
							\pm
							2.8
						$
					\\
			  All
			& $1.93$%
						& $
							134.2
							\pm
							19.3
							\pm
							2.2
						$
						& $
							164.4
							\pm
							19.9
							\pm
							2.7
						$
						& $
							202.3
							\pm
							14.1
							\pm
							4.9
						$
						& $
							267.8
							\pm
							10.2
							\pm
							1.7
						$
					\\
			  All
			& $2.06$%
						& $
							137.7
							\pm
							18.6
							\pm
							1.9
						$
						& $
							196.0
							\pm
							16.3
							\pm
							0.7
						$
						& $
							211.5
							\pm
							13.2
							\pm
							0.2
						$
						& $
							242.3
							\pm
							11.0
							\pm
							4.5
						$
					\\ \enddata
 \tablecomments{
 The circular velocities $V_c(R)$ inferred from weak lensing for four baryonic mass bins with bin edges $\log_{10} M_b/M_\odot = [9.0, 10.5, 10.8, 11.1, 11.5]$.
 We separately list results for ETGs, LTGs, and the entire sample.
 The lowest mass bin for ETGs is not shown because it contains too few lenses to obtain a useful signal.
 The listed errors on $V_c$ are the statistical errors (see the colored error bars in Figure~\ref{fig:vc}) and the systematic errors from converting  ESD profiles to radial accelerations (see Section~\ref{sec:method:vc}).
 Covariance matrices are available on reasonable request to the authors.
 Measurements are omitted for radial bins with negative stacked radial acceleration.
 Table~\ref{tab:vc} is published in its entirety in the machine-readable format.
 A portion is shown here for guidance regarding its form and content.
 }
\end{deluxetable*}

The circular velocity curves of NFW halos do not remain flat indefinitely.
This is in tension with the lensing-inferred circular velocities that remain flat out to $\sim 1\,\mathrm{Mpc}$.
In the higher-mass bins, there is no strong discrepancy if we conservatively consider only $R \lesssim 300\,\mathrm{kpc}$ where there is no doubt about the isolation criterion.
In the lower-mass bins, however, there is a clear discrepancy already at $R\lesssim 300$ kpc because of the smaller DM halos.
The lensing data now probe to the virial radius and beyond with no indication of the expected downturn in rotation speed.

Although we here assume a specific DM halo profile, our results apply more generally in the context of $\Lambda$CDM cosmology because our lensing data mainly probe the outer slopes of DM halos which are predicted to approximately follow $\rho_{\rm DM} \propto r^{-3}$ due to the hierarchical process of structure formation. The gravitational effect of baryons is expected to lead to halo contraction, which makes the circular velocities decline faster than for a fiducial NFW profile \citep{Li2022a, Li2022b}. Other baryonic processes, such as stellar and black-hole feedback,
play a negligible role for the outer halo slope. Thus, assuming different reasonable DM halo profiles, such as those from hydrodynamic simulations of galaxy formation \citep[see][]{Li2020}, will not affect our conclusions.
Similarly, different stellar mass-halo mass or mass-concentration relations do not change our conclusions because these mainly change the overall normalization of $V_c$ but leave its shape at large radii unchanged.

Our isolation criterion may preferentially select void galaxies that, in principle, may be hosted by systematically different halos than the average galaxy. In practice, however, variations in halo concentration with cosmic environment are predicted to be small \citep{Hellwing2021}. If anything, they lead to slightly larger concentrations for void galaxies in the relevant range of halo masses ($>10^{11}\,M_\odot$), which would make the circular velocities decline even earlier than shown in Figure~\ref{fig:vc-sparc}.

No galaxy is completely isolated. To illustrate the possible effect of neighboring DM halos, the light green band in Figure~\ref{fig:vc-sparc} shows a qualitative estimate (see Appendix~\ref{sec:appendix:twohalo}) of the so-called two-halo term. The approach we employ should not be considered the immutable prediction of $\Lambda$CDM, but it does successfully reproduce the lensing signal at large radii around isolated lenses in $\Lambda$CDM simulations (Appendix~\ref{sec:appendix:twohalo}).

\subsection{BTFR}
\label{sec:results:btfr}

\begin{figure*}
\plotone{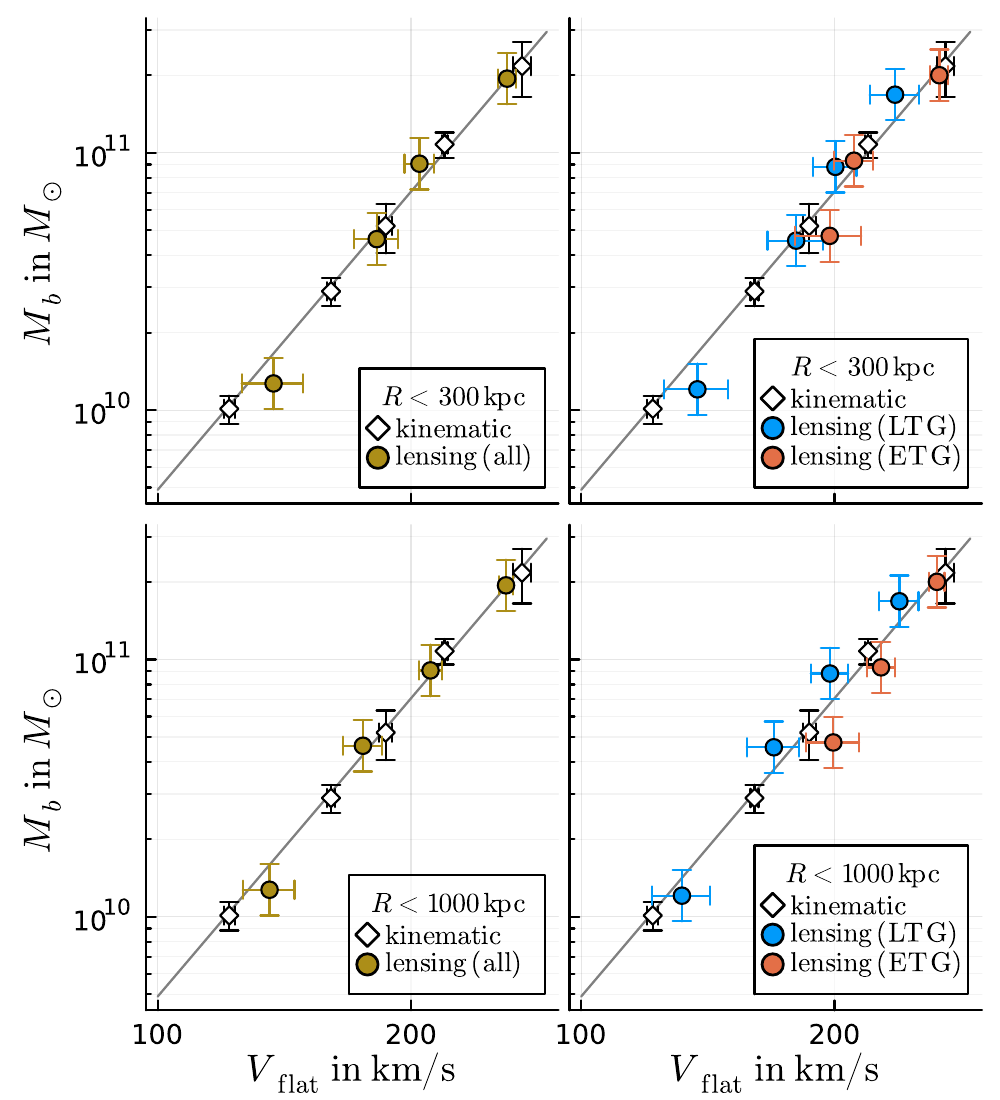}
 \caption{
   The baryonic Tully-Fisher relation implied by weak lensing for the entire sample (yellow symbols, left column) and for ETGs and LTGs separately (red and blue symbols, right column).
   The $V_{\mathrm{flat}}$ values are weighted averages of the $V_c$ values shown in Figure~\ref{fig:vc} for $50\,\mathrm{kpc} < R < 300\,\mathrm{kpc}$ (first row) and $50\,\mathrm{kpc} < R < 1000\,\mathrm{kpc}$ (second row).
   Vertical error bars represent a $0.1\,\mathrm{dex}$ systematic uncertainty on $M_\ast/L$.
   For comparison, we also show the best fit to the kinematic data from \citet{Lelli2019} (solid gray line) and the corresponding binned kinematic data (white diamonds).
 }
 \label{fig:btfr}
\end{figure*}

Figure~\ref{fig:btfr} and Table~\ref{tab:btfr} show the BTFR from weak lensing data and compare it with binned kinematic data from \citet{Lelli2019}.
For visualization purposes, the mass bins for the kinematic data are chosen such that the data points do not overlap with those from weak lensing.
The kinematic $V_{\mathrm{flat}}$ values are averaged in each mass bin, with weights proportional to their inverse squared uncertainties.
As with the weak-lensing data, we use properly averaged baryonic masses (see Section~\ref{sec:method:vflat}).

We consider $V_{\mathrm{flat}}$ to be the average of the data points over two ranges of radii weighted by their statistical uncertainty (see Section~\ref{sec:method:vflat}).
Our results are most robust over the radial range $50\,\mathrm{kpc} < R < 300\,\mathrm{kpc}$, so Figure~\ref{fig:btfr} shows the result of averaging over these data. 
We also show the result from averaging over $50\,\mathrm{kpc} < R < 1000\,\mathrm{kpc}$. 
These give similar results; we have checked other reasonable choices of radii to include in the computation of $V_{\mathrm{flat}}$ and have verified that differences between them are generally smaller than the statistical uncertainties.

Figure~\ref{fig:btfr} shows good agreement between the BTFR from lensing data and that known from kinematics. 
The lensing data are consistent with being an extension of the flat rotation curves that are observed kinematically. This extension persists indefinitely in radius at the amplitude indicated by the baryonic mass.

When considering the entire lensing sample, there is no hint of an offset between the kinematic and lensing BTFR.
When dividing by galaxy type, there is the suggestion of a small offset between ETGs and LTGs, but it is not statistically significant.
Based on the random uncertainties in $V_{\mathrm{flat}}$ alone, the difference corresponds to $0.54\sigma$ and $1.86\sigma$ when $V_{\mathrm{flat}}$ is defined using radii up to $300\,\mathrm{kpc}$ and $1000\,\mathrm{kpc}$, respectively. 
The latter is also subject to type-dependent differences in the isolation criterion, which is more robust for LTGs than for ETGs \citep{Mistele2023d}. The net effect of this systematic difference would be a slight overestimate of $V_{\mathrm{flat}}$ for ETGs when including larger radii.

\begin{deluxetable}{lRR}
 \caption{Baryonic Tully-Fisher Data \label{tab:btfr}}
 \tablehead{
		  \colhead{Sample}
		& \colhead{$\log_{10} M_{b}$}
		& \colhead{$V_{\mathrm{flat}}$}%
		\\
		  \colhead{}
		& \colhead{$M_\odot$}
 		& \colhead{$\mathrm{km/s}$}}\startdata	%
			       All\;($R < 300\,\mathrm{kpc}$)
			    & $
					10.10
					\phn\phn\phd
					\phn\phd\phn\phn
				  $
				& $
					137.3
					\pm
					11.5
				  $\\%
			       All\;($R < 300\,\mathrm{kpc}$)
			    & $
					10.66
					\phn\phn\phd
					\phn\phd\phn\phn
				  $
				& $
					182.1
					\pm
					10.8
				  $\\%
			       All\;($R < 300\,\mathrm{kpc}$)
			    & $
					10.96
					\phn\phn\phd
					\phn\phd\phn\phn
				  $
				& $
					204.7
					\pm
					8.2
					\phn%
				  $\\%
			       All\;($R < 300\,\mathrm{kpc}$)
			    & $
					11.29
					\phn\phn\phd
					\phn\phd\phn\phn
				  $
				& $
					260.2
					\pm
					6.2
					\phn%
				  $ \\ \hline %
			       LTG\;($R < 300\,\mathrm{kpc}$)
			    & $
					10.08
					\phn\phn\phd
					\phn\phd\phn\phn
				  $
				& $
					137.4
					\pm
					11.9
				  $\\%
			       LTG\;($R < 300\,\mathrm{kpc}$)
			    & $
					10.66
					\phn\phn\phd
					\phn\phd\phn\phn
				  $
				& $
					180.1
					\pm
					13.6
				  $\\%
			       LTG\;($R < 300\,\mathrm{kpc}$)
			    & $
					10.95
					\phn\phn\phd
					\phn\phd\phn\phn
				  $
				& $
					200.5
					\pm
					11.9
				  $\\%
			       LTG\;($R < 300\,\mathrm{kpc}$)
			    & $
					11.23
					\phn\phn\phd
					\phn\phd\phn\phn
				  $
				& $
					236.1
					\pm
					15.9
				  $ \\ \hline %
			       ETG\;($R < 300\,\mathrm{kpc}$)
			    & $
					10.68
					\phn\phn\phd
					\phn\phd\phn\phn
				  $
				& $
					197.4
					\pm
					17.7
				  $\\%
			       ETG\;($R < 300\,\mathrm{kpc}$)
			    & $
					10.97
					\phn\phn\phd
					\phn\phd\phn\phn
				  $
				& $
					211.0
					\pm
					11.3
				  $\\%
			       ETG\;($R < 300\,\mathrm{kpc}$)
			    & $
					11.30
					\phn\phn\phd
					\phn\phd\phn\phn
				  $
				& $
					266.5
					\pm
					6.7
					\phn%
				  $ \\ \hline \tablehead{
		  \colhead{Sample}
		& \colhead{$\log_{10} M_{b}$}
		& \colhead{$V_{\mathrm{flat}}$}%
		\\
		  \colhead{}
		& \colhead{$M_\odot$}
 		& \colhead{$\mathrm{km/s}$}}%
			       All\;($R < 1000\,\mathrm{kpc}$)
			    & $
					10.11
					\phn\phn\phd
					\phn\phd\phn\phn
				  $
				& $
					135.8
					\pm
					9.6
					\phn%
				  $\\%
			       All\;($R < 1000\,\mathrm{kpc}$)
			    & $
					10.66
					\phn\phn\phd
					\phn\phd\phn\phn
				  $
				& $
					175.3
					\pm
					9.3
					\phn%
				  $\\%
			       All\;($R < 1000\,\mathrm{kpc}$)
			    & $
					10.96
					\phn\phn\phd
					\phn\phd\phn\phn
				  $
				& $
					211.0
					\pm
					6.6
					\phn%
				  $\\%
			       All\;($R < 1000\,\mathrm{kpc}$)
			    & $
					11.29
					\phn\phn\phd
					\phn\phd\phn\phn
				  $
				& $
					259.4
					\pm
					5.1
					\phn%
				  $ \\ \hline %
			       LTG\;($R < 1000\,\mathrm{kpc}$)
			    & $
					10.08
					\phn\phn\phd
					\phn\phd\phn\phn
				  $
				& $
					131.8
					\pm
					10.4
				  $\\%
			       LTG\;($R < 1000\,\mathrm{kpc}$)
			    & $
					10.66
					\phn\phn\phd
					\phn\phd\phn\phn
				  $
				& $
					169.4
					\pm
					12.1
				  $\\%
			       LTG\;($R < 1000\,\mathrm{kpc}$)
			    & $
					10.95
					\phn\phn\phd
					\phn\phd\phn\phn
				  $
				& $
					197.5
					\pm
					10.1
				  $\\%
			       LTG\;($R < 1000\,\mathrm{kpc}$)
			    & $
					11.23
					\phn\phn\phd
					\phn\phd\phn\phn
				  $
				& $
					238.9
					\pm
					12.8
				  $ \\ \hline %
			       ETG\;($R < 1000\,\mathrm{kpc}$)
			    & $
					10.68
					\phn\phn\phd
					\phn\phd\phn\phn
				  $
				& $
					199.4
					\pm
					14.5
				  $\\%
			       ETG\;($R < 1000\,\mathrm{kpc}$)
			    & $
					10.97
					\phn\phn\phd
					\phn\phd\phn\phn
				  $
				& $
					227.2
					\pm
					8.7
					\phn%
				  $\\%
			       ETG\;($R < 1000\,\mathrm{kpc}$)
			    & $
					11.30
					\phn\phn\phd
					\phn\phd\phn\phn
				  $
				& $
					264.7
					\pm
					5.6
					\phn%
				  $ \\ \hline %
				      Kinematic
				    & $
						8.69
						\pm
						0.06
					  $
					& $
						54.2
						\pm
						1.0
						\phn
					  $\\%
				      Kinematic
				    & $
						9.25
						\pm
						0.04
					  $
					& $
						77.9
						\pm
						1.1
						\phn
					  $\\%
				      Kinematic
				    & $
						9.67
						\pm
						0.08
					  $
					& $
						89.1
						\pm
						1.3
						\phn
					  $\\%
				      Kinematic
				    & $
						10.00
						\pm
						0.05
					  $
					& $
						121.7
						\pm
						1.8
						\phn
					  $\\%
				      Kinematic
				    & $
						10.46
						\pm
						0.05
					  $
					& $
						160.7
						\pm
						1.7
						\phn
					  $\\%
				      Kinematic
				    & $
						10.72
						\pm
						0.09
					  $
					& $
						186.6
						\pm
						3.3
						\phn
					  $\\%
				      Kinematic
				    & $
						11.03
						\pm
						0.05
					  $
					& $
						219.3
						\pm
						1.8
						\phn
					  $\\%
				      Kinematic
				    & $
						11.34
						\pm
						0.10
					  $
					& $
						271.0
						\pm
						6.4
						\phn
					  $ \enddata
 \tablecomments{
  The $V_{\mathrm{flat}}$ values and the corresponding properly-averaged baryonic masses used for the BTFR (see Figure~\ref{fig:btfr}).
  The $V_{\mathrm{flat}}$ errors listed are the statistical errors.
  The uncertainty on the mean $M_b$ is statistically negligible for the lensing data, but we adopt $0.1\,\mathrm{dex}$ as a systematic uncertainty in $M_\ast/L$, see Figure~\ref{fig:btfr}.
 }
\end{deluxetable}

\section{Discussion}

The circular velocity curves from weak lensing observations remain flat for hundreds of kpc, possibly up to $1\,\mathrm{Mpc}$, and imply a weak-lensing BTFR that is fully consistent with the kinematic BTFR.
These results apply to both LTGs and ETGs separately.
Evidently, the asymptotic flatness of rotation curves and the BTFR are independent of galaxy morphology (disks or spheroids) and evolutionary history (star-forming or passive).
Galaxies seem inevitably to lie on the BTFR given the availability of an adequate tracer to measure $V_c$ out to large radii. Similar results, indeed, were found for ETGs that occasionally possess an outer extended HI disk \citep{denHeijer2015,Shelest2020}.

The mere existence of the BTFR \citep{McGaugh2000} already suggests that this relation is independent of the diverse evolutionary histories of galaxies because the relative contribution of gas mass and stellar mass can greatly vary across the galaxy population \citep[e.g.,][]{Lelli2022}.
In general, the gas mass contribution becomes important below $\log_{10} M_b/M_{\odot} = 10$ \citep{McGaugh2000}, so it should not matter much for our lens sample that contains mostly galaxies with $\log_{10} M_b/M_{\odot} > 10$ \citep{Mistele2023d}. Indeed, we have verified that rerunning our analysis using stellar masses instead of baryonic masses has only a small effect on the resulting Tully-Fisher relation, with the data points in Figure~\ref{fig:btfr} moving towards slightly smaller masses. This is expected and implies that the details of our gas mass estimates from Equations~\eqref{eq:fhot} and~\eqref{eq:fcold} are relatively unimportant to our general results.

Given that kinematic measurements of $V_{\mathrm{flat}}$ rely on data from relatively small radii where the disk geometry matters, rotation curves may have not yet reached their truly asymptotic value.
Consequently, one may expect that lensing data gives slightly smaller $V_{\mathrm{flat}}$ values compared to the kinematic BTFR.
Indeed, a razor-thin disk galaxy with scale length $5\,\mathrm{kpc}$ that obeys the radial acceleration relation \citep{Lelli2017b,Brouwer2021,Mistele2023d} has an asymptotic $V_{\mathrm{flat}}$ that is about $5\%$ smaller than the rotation curve velocity at $25\,\mathrm{kpc}$ \citep{McGaugh1998b}.
There are hints of just such an offset in the higher LTG mass bins, but in general the uncertainties in our weak-lensing analysis do not allow for such a small offset to be reliably observed.

Our results are difficult to understand in $\Lambda$CDM because the data should have reached the asymptotic $V_c \sim \sqrt{\log(r)/r}$ decline of the DM halos.
On the other hand, a universal BTFR and indefinitely flat rotation curves for isolated galaxies were predicted by Modified Newtonian Dynamics \citep[MOND,][]{Milgrom1983a, Milgrom1983b, Milgrom1983c}.
Indeed, the lensing data now probe far into the deep MOND regime of extremely low accelerations without showing any deviation from the prediction that the asymptotic $V_{\mathrm{flat}}$ is determined by $M_b$ as $V_{\mathrm{flat}} \propto M_b^{1/4}$.
Perhaps with further improvement in the data it might become possible to perceive a decline at large radii due to the so-called external field effect \citep{Bekenstein1984, Chae2020, Chae2021}.
Given our strict isolation criterion and the stacking required to obtain the lensing signal, this external field effect may be undetectable in the present data.

\section{Conclusion}

We have derived circular velocities for isolated galaxies from weak gravitational lensing data.
The circular velocity curves are consistent with being flat out to hundreds of $\mathrm{kpc}$, perhaps even $1\,\mathrm{Mpc}$, with no sign of having reached the edge of the DM halo.
Using these circular velocities, we have constructed the BTFR implied by weak lensing, finding good agreement with previous kinematic determinations of the BTFR.
These results hold for both LTGs and ETGs separately, suggesting a common universal behavior.

\begin{acknowledgments}
TM thanks Amel Durakovic for helpful discussions.
This work was supported by the DFG (German Research Foundation) – 514562826.
Based on observations made with ESO Telescopes at the La Silla Paranal Observatory under programme IDs 177.A-3016, 177.A-3017, 177.A-3018 and 179.A-2004, and on data products produced by the KiDS consortium. The KiDS production team acknowledges support from: Deutsche Forschungsgemeinschaft, ERC, NOVA and NWO-M grants; Target; the University of Padova, and the University Federico II (Naples).
GAMA is a joint European-Australasian project based around a spectroscopic campaign using the Anglo-Australian Telescope. The GAMA input catalogue is based on data taken from the Sloan Digital Sky Survey and the UKIRT Infrared Deep Sky Survey. Complementary imaging of the GAMA regions is being obtained by a number of independent survey programmes including GALEX MIS, VST KiDS, VISTA VIKING, WISE, Herschel-ATLAS, GMRT and ASKAP providing UV to radio coverage. GAMA is funded by the STFC (UK), the ARC (Australia), the AAO, and the participating institutions. The GAMA website is https://www.gama-survey.org/~. 

\end{acknowledgments}

\clearpage %
\appendix

\section{BTFR from stacking}
\label{sec:appendix:btfr}

In Section~\ref{sec:method:vc} we define stacked circular velocities $V_c$ that we obtain from stacked radial accelerations $g_{\mathrm{obs}}^{\mathrm{stacked}}(R)$ as $V_c(R) \equiv (R \, g_{\mathrm{obs}}^{\mathrm{stacked}}(R))^{1/2}$.
These circular velocities are not linear averages of the circular velocities of the individual stacked lens galaxies, i.e. they are not $\langle V_c \rangle$.
Instead, they are $\sqrt{\langle V_c^2 \rangle}$.
In Section~\ref{sec:method:vflat}, we average these circular velocities at different radii to obtain a flat circular velocity $V_{\mathrm{flat}}$ to be used when constructing the weak-lensing BTFR.
Since the BTFR relation $V_{\mathrm{flat}} \propto M_b^{1/4}$ concerns individual galaxies and since our $V_{\mathrm{flat}}$ is obtained using a somewhat involved stacking and averaging procedure, we should average and stack the baryonic mass to compare $V_{\mathrm{flat}}$ to in a similar way.
That is, in order to test the BTFR, we should use a definition of the averaged baryonic mass such that, if the relation $V_{\mathrm{flat}} \propto M_b^{1/4}$ holds for individual galaxies it also holds for the stacked and averaged $V_{\mathrm{flat}}$ we use.

The correctly stacked and averaged baryonic masses corresponding to our stacked and averaged $V_{\mathrm{flat}}$ are obtained as follows.
First, we calculate a properly stacked and weighted $\sqrt{M_b}$ for each radial bin.
Using our notation from Section~\ref{sec:method:vc},
\begin{align}
 \label{eq:avgsqrtMb}
 \langle \sqrt{M_b} \rangle(R) \equiv \bar{N}^{-1}(R) \sum_l \bar{w}_l(R) \sqrt{M_{b,l}} \,.
\end{align}
This mirrors how we obtain the stacked and weighted $\langle V_c^2 \rangle$ and follows the relation $V_{\mathrm{flat}}^2 \propto \sqrt{M_b}$ for individual galaxies.
Then, mirroring how we go from $\langle V_c^2 \rangle(R)$ to $\sqrt{\langle V_c^2 \rangle(R)}$, we take the square root of Equation~\eqref{eq:avgsqrtMb}.
Finally, we take the weighted average of different radii, mirroring how we obtain $V_{\mathrm{flat}}$ from $\sqrt{\langle V_c \rangle(R)}$ in Section~\ref{sec:method:vflat}.
This gives the effective averaged baryonic mass $M_{b,\mathrm{eff}}$ to be used in our weak-lensing BTFR,
\begin{align}
 M_{b,\mathrm{eff}} \equiv \left( \tilde{N}^{-1} \sum_R \tilde{w}(R) \sqrt{\langle \sqrt{M_b} \rangle (R) } \right)^4 \,.
\end{align}
Here, the sum goes over the radial bins that we average over, the weights $\tilde{w}(R)$ are given by $\sigma_{V_c}^{-2}(R)$ where $\sigma_{V_c}$ is the statistical uncertainty of $V_c$ and $\tilde{N} = \sum_R \tilde{w}(R)$ normalizes the weights.
This definition of the averaged baryonic mass has the desired property that, if the relation $V_{\mathrm{flat}} \propto M_b^{1/4}$ holds for individual galaxies, it also holds for the stacked and averaged $V_{\mathrm{flat}}$ and $M_{b,\mathrm{eff}}$ we use here.

\section{Statistical uncertainties}
\label{sec:appendix:staterrors}

To derive an accurate expression for the statistical uncertainty of the deprojected radial acceleration $g_{\mathrm{obs}}$ from Equation~\eqref{eq:gobs_from_esd}, we consider a discretized version of the integral in Equation~\eqref{eq:gobs_from_esd}.
This discretized version is what we evaluate numerically to obtain $g_{\mathrm{obs}}$. 
Specifically, Equation~\eqref{eq:gobs_from_esd} can be written in the form
\begin{align}
 \label{eq:gobs_from_esd_discretized}
 g_{\mathrm{obs}}(R_\alpha) = 4G \sum_{i=\alpha}^N C_{\alpha i} \Delta \Sigma_i \,.
\end{align}
To obtain this form, 
following \citet{Mistele2023d}, we linearly interpolate $\Delta \Sigma$ between the discrete radial bins where it is measured.
Here, $R_1, R_2, \dots, R_N$ denote the bin centers of the radial bins where $\Delta \Sigma$ is measured in increasing order, $\alpha$ and $i \geq \alpha$ each indicate one of these $N$ bins, and $\Delta \Sigma_i$ is the value of $\Delta \Sigma$ in bin $i$.
The coefficients $C_{\alpha i}$ are constants that are independent of the measured values $\Delta \Sigma_i$ (see below for their definition).
Equation~\eqref{eq:gobs_from_esd_discretized} follows by splitting the integral in Equation~\eqref{eq:gobs_from_esd} at the bin centers $R_i$ and analytically evaluating the integral in each bin, using the fact that we linearly interpolate between the $\Delta \Sigma_i$ data points.

The statistical uncertainty on $g_{\mathrm{obs}}(R_\alpha)$ and the covariances between the different radial bins can be directly read off from Equation~\eqref{eq:gobs_from_esd_discretized}.
In particular, following \citet{Mistele2023d} and \citet{Brouwer2021} (see also \citet{Viola2015}), we consider the statistical uncertainty from the ellipticities of the source galaxies.
Thus, for an individual lens, the $\Delta \Sigma_i$ in different radial bins are uncorrelated and we have the covariance
\begin{align}
\begin{split}
\label{eq:gobs_cov_individual}
\Cov(g_{\mathrm{obs}}(R_\alpha), g_{\mathrm{obs}}(R_\beta))
&= (4G)^2 \sum_{i = \alpha}^N \sum_{j = \beta}^N C_{\alpha i} C_{\beta j} \Cov(\Delta \Sigma_i, \Delta \Sigma_j) \\
&= (4G)^2 \sum_{i = \max(\alpha, \beta)}^N C_{\alpha i} C_{\beta i} \sigma^2_{\Delta \Sigma_i} \,,
\end{split}
\end{align}
where $\sigma_{\Delta \Sigma_i}$ is the statistical uncertainty on $\Delta \Sigma_i$ that we calculate as in \citet{Mistele2023d}.

As discussed in \citet{Mistele2023d}, the radial accelerations $g_{\mathrm{obs},l}(R_\alpha)$ and $g_{\mathrm{obs},l'}(R_\beta)$ of two different lenses $l$ and $l'$ are to a good approximation uncorrelated.
This is because our lens galaxies are isolated so that the source galaxies only rarely contribute to multiple lenses simultaneously.
Thus, for the stacked radial acceleration from Equation~\eqref{eq:gobs_stacked}, we have to a good approximation
\begin{align}
\begin{split}
\Cov(g_{\mathrm{obs}}^{\mathrm{stacked}}(R_\alpha), g_{\mathrm{obs}}^{\mathrm{stacked}}(R_\beta))
&= \bar{N}^{-1}(R_\alpha) \bar{N}^{-1}(R_\beta) \sum_l \bar{w}_l(R_\alpha) \bar{w}_{l}(R_\beta) \Cov(g_{\mathrm{obs},l}(R_\alpha), g_{\mathrm{obs},l}(R_\beta)) \,,
\end{split}
\end{align}
where the covariance of the lens $l$ on the right-hand side is to be calculated as in Equation~\eqref{eq:gobs_cov_individual}.
The diagonal entries of this covariance matrix are the squared statistical uncertainties $\sigma_{g_{\mathrm{obs}}^{\mathrm{stacked}}}^2$.
We calculate the covariance matrices and statistical uncertainties for the circular velocities $V_c = (R g_{\mathrm{obs}}^{\mathrm{stacked}})^{1/2}$ and the radially-averaged flat circular velocity $V_{\mathrm{flat}}$ using linear error propagation.

It remains to give the definition of $C_{\alpha i}$, which is
\begin{align}
 C_{\alpha i}      &\equiv \begin{cases}
                            \Delta \theta_{\alpha \alpha} - f_{\alpha \alpha} \,, & \mathrm{for}\;\alpha = i < N \\
                            \Delta \theta_{\alpha i} - f_{\alpha i} + f_{\alpha,i-1} \,, & \mathrm{for}\;\alpha < i < N \\
                            f^{\mathrm{cont}}_{\alpha N} + f_{\alpha,N-1},  & \mathrm{for}\; \alpha < i =  N \\
                            f^{\mathrm{cont}}_{N N} , & \mathrm{for}\; \alpha = i = N
                           \end{cases} \,,
\end{align}
where, for $\alpha \leq i < N$, we further define
\begin{align}
 \Delta \theta_{\alpha i} \equiv \theta_{\alpha i} - \theta_{\alpha,i+1} \equiv \arcsin\left(\frac{R_\alpha}{R_i}\right) - \arcsin\left(\frac{R_{\alpha}}{R_{i+1}}\right) \,,
\end{align}
and
\begin{align}
 f_{\alpha i} \equiv \frac{
  -R_\alpha \arctanh\left(\sqrt{1 - \left(\frac{R_\alpha}{R_i}\right)^2}\right)
  +R_\alpha \arctanh\left(\sqrt{1 - \left(\frac{R_\alpha}{R_{i+1}}\right)^2}\right)
  - R_i \Delta \theta_{\alpha i}
 }{
  R_{i+1} - R_i
 } \,.
\end{align}
The $f^{\mathrm{cont}}_{\alpha N}$ encode how we extrapolate $\Delta \Sigma$ beyond the last measured data point.
Following \citet{Mistele2023d}, we assume $\Delta \Sigma \propto 1/R$ there, which corresponds to a singular isothermal sphere.
The uncertainty in this choice is taken into account as a systematic error as described in Section~\ref{sec:method:vc}. 
We have
\begin{align}
 \left. f^{\mathrm{cont}}_{\alpha N} \right|_{\mathrm{SIS\,extrapolation}} = \frac{R_N}{R_\alpha} \left(1 - \sqrt{1 - \left(  \frac{R_\alpha}{R_N} \right)^2} \right) \,.
\end{align}

\section{Qualitative two-halo term estimate}
\label{sec:appendix:twohalo}

Consider the contribution $\Delta \Sigma_e$ of a lens' environment to the observed  ESD profile $\Delta \Sigma$.
A simple estimate of $\Delta \Sigma_e$ is given by \citep[e.g.][]{Guzik2001,Oguri2011,Covone2014}
\begin{align}
\label{eq:esd_twohalo}
 \Delta \Sigma_e (R) = b_e \frac{\bar{\rho}_{m,0}}{2 \pi D_l^2} \int_0^\infty d \ell \, \ell J_2\left(\frac{\ell R}{D_l}\right) P_m(k_\ell; z) \,,
\end{align}
where $J_2$ denotes the second Bessel function of the first kind, $D_l$ is the angular diameter distance of the lens, $b_e$ is the bias, $z$ is the redshift of the lens, $\bar{\rho}_{m,0}$ is the mean matter density at redshift $z=0$, and $P_m(k_\ell; z)$ is the linear matter power spectrum at $k_\ell = \ell/[(1+z) D_l]$ (see below for why we use the \emph{linear} power spectrum).
This leads to an additional contribution $g_{\mathrm{obs},e}$ to the acceleration $g_{\mathrm{obs}}$ that we infer when using Equation~\eqref{eq:gobs_from_esd} 
\begin{align}
 \label{eq:gobs_twohalo}
g_{\mathrm{obs},e}(R)
= 4G \int_0^{\pi/2} d \theta \, \Delta \Sigma_e \left(\frac{R}{\sin \theta}\right) 
= \frac{4G}{R} \cdot b_e \frac{\bar{\rho}_{m,0}}{2 \pi D_l} \int_0^\infty d\ell P_m(k_\ell; z) \left(\frac{D_l }{\ell R} \sin\left(\frac{\ell R}{D_l}\right) - \cos\left(\frac{\ell R}{D_l}\right) \right) \,.
\end{align}
The second equality follows by exchanging the order of the $\theta$ and $\ell$ integrals, substituting $x \equiv (\ell R)/(D_l \sin \theta)$ at fixed $\ell$ in the inner $\theta$ integral, and analytically evaluating the resulting $x$ integral using Mathematica \citep{Mathematica14}.
We evaluate the remaining $\ell$ integral numerically, assuming $z=0.2$ for a typical lens galaxy.
We calculate the linear matter power spectrum using CAMB \citep{Lewis2002}.

\begin{figure*}
\plotone{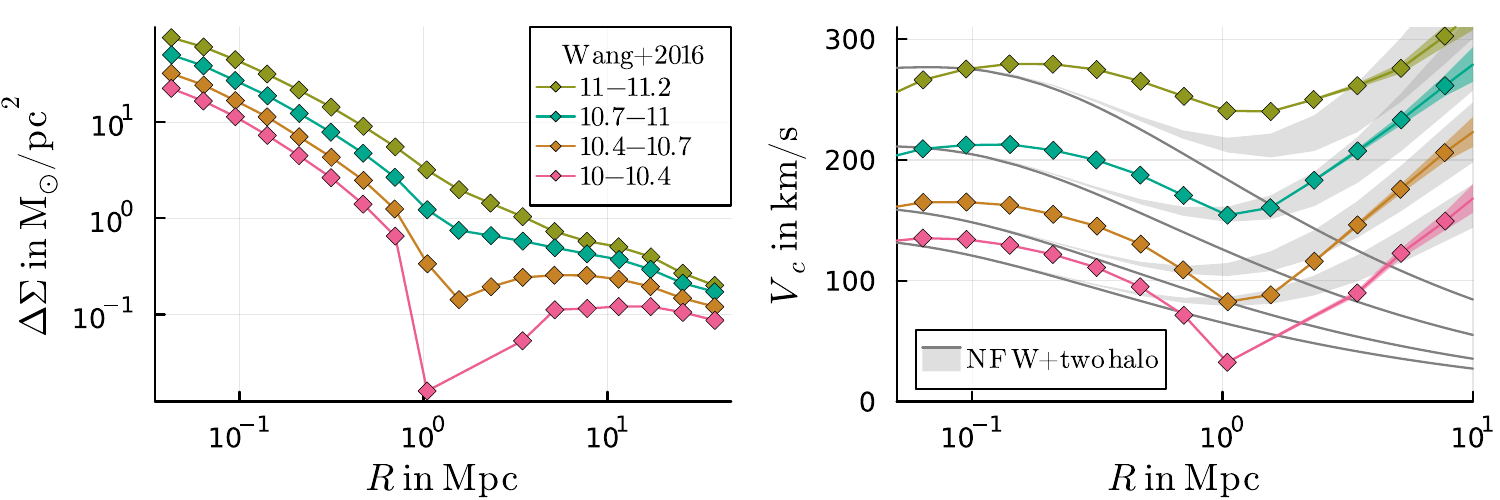}
 \caption{
 Left:
 Stacked ESD profiles for four stellar mass bins from the G-11P' $\Lambda$CDM simulation from \citet{Wang2016} (see their Figure~4).
 We ignore data points with negative $\Delta \Sigma$ in the lowest mass bin.
 Right: The circular velocities implied by these ESD profiles according to our deprojection formula Equation~\eqref{eq:gobs_from_esd}.
 Colored bands indicate the systematic uncertainty from converting ESD profiles to accelerations (Section~\ref{sec:method:vc}).
 Solid gray lines indicate the circular velocity of an NFW halo and a baryonic point mass as in Figure~\ref{fig:vc-sparc}.
 Light gray bands additionally take into account contributions due to the local environment around not perfectly isolated lenses according to Equation~\eqref{eq:gobs_twohalo}.
 In particular, the gray bands show the $\pm0.1\,\mathrm{dex}$ range around $b_e$ values of $0.3, 0.57, 0.95, 1.25$, respectively, for the four stellar mass bins. 
 }
 \label{fig:twohalo}
\end{figure*}

One limitation is that Equation~\eqref{eq:esd_twohalo} does not know that our lens sample consists of isolated galaxies.
For simplicity, we here assume that this can be taken into account by choosing an appropriate normalization $b_e$ and using the linear, rather than non-linear, matter power spectrum $P_m$.
Indeed, using the linear power spectrum qualitatively mimics the effect of our isolation criterion of removing structure on relatively small spatial scales.
To find reasonable values for $b_e$ we match our semi-analytical estimate to the lensing signal of isolated galaxies obtained from $\Lambda$CDM simulations by \citet{Wang2016}.
Since \citet{Wang2016} impose an isolation criterion different from ours (see below), this amounts to making the additional assumption that, at least qualitatively, the $b_e$ values we obtain in this way apply to our isolation criterion as well.

We use stacked ESD profiles for four stellar mass bins from the G11-P' model from \citet{Wang2016}.
These cover a similar range in $V_c$ as our KiDS data.
Figure~\ref{fig:twohalo}, left, shows these ESD profiles and Figure~\ref{fig:twohalo}, right, shows the corresponding $V_c$ inferred using our deprojection formula Equation~\eqref{eq:gobs_from_esd}.
The inferred circular velocities $V_c$ show an upwards trend at large radii.
This is due to the two-halo term which leads to $\Delta \Sigma$ falling off slower than $1/R$ which in turn means our deprojection formula Equation~\eqref{eq:gobs_from_esd} infers rising circular velocities.
We note that, where the two-halo term becomes important, these inferred circular velocities no longer correspond to actual circular orbits of bound objects.
They are simply what our method infers when applied to non-isolated lenses.

This behavior can, at least qualitatively, be reproduced by considering the environment-induced lensing-inferred acceleration $g_{\mathrm{obs},e}$ from Equation~\eqref{eq:gobs_twohalo} in addition to an NFW halo and a baryonic point mass.
This is illustrated in Figure~\ref{fig:twohalo}, right, where we show a $\pm 0.1\,\mathrm{dex}$ band around $b_e$ values of $0.3$, $0.57$, $0.95$, and $1.25$, respectively, for the four stellar mass bins of \citet{Wang2016}.
These values are relatively small \citep{Tinker2010} as might be expected for a sample of isolated galaxies.
We also see a clear overall trend with lower masses requiring lower values of $b_e$.
As a rough approximation, we show these same four bands of $b_e$ values in Figure~\ref{fig:vc-sparc} in the main text. 

The NFW halo plus baryonic point mass models shown in Figure~\ref{fig:twohalo} are calculated as in Section~\ref{sec:results:vc} but with $\log_{10} M_\ast/M_\odot$ values $10.2$, $10.5$, $10.85$, and $11.1$, respectively, for the four stellar mass bins.
Baryonic masses are calculated using these stellar masses assuming the cold gas mass estimate from Equation~\eqref{eq:fcold} for the three lower mass bins and the hot gas mass estimate from Equation~\eqref{eq:fhot} for the highest mass bin.
These values were chosen to fit the overall normalization of the inferred circular velocities and may not be correct in detail.
For our purposes, the important point is the tail where the two-halo term dominates. 
The details of the NFW halos and baryonic point masses are unimportant.

Nevertheless, one may notice from Figure~\ref{fig:twohalo}, right, that our NFW plus baryonic point mass models do not match the shape of the circular velocities inferred from the \citet{Wang2016} ESD profiles, not even at small radii where the two-halo term should be unimportant.
We speculate that this may be due the particular isolation criterion imposed by \citet{Wang2016}.
Indeed, one difference compared to our isolation criterion (Section~\ref{sec:data}) is that \citet{Wang2016} only exclude lenses with neighbors that are brighter than the lens itself, i.e. $f_\ast = 1.0$ using our notation from Section~\ref{sec:data} and assuming a constant $M_\ast/L$.
This allows for significant non-isolation effects already at small radii and may be why NFW plus baryonic point mass models are not a good match.
In any case, as mentioned above, the details at small radii do not matter for our purposes, which is merely to illustrate the effects of the two-halo term.

\bibliography{lensing-BTFR}{}
\bibliographystyle{aasjournal}

\end{document}